\documentclass[12pt,pdfstartview=FitH]{article}

\usepackage{amssymb}
\usepackage{amsmath}
\usepackage{bbm}
\usepackage{enumitem}
\usepackage{physics}
\usepackage{slashed}
\usepackage{xcolor}
\usepackage{jheppub}
\usepackage{tikz}
\usepackage{empheq}
\usepackage{picture}
\usepackage[Smaller,makeroom,thicklines]{cancel}

\title{Exact thermal correlators at vanishing momenta and the black hole singularity}
\author{Kaustubh Singhi}
\affiliation{International Centre for Theoretical Sciences (ICTS-TIFR), \\ Tata Institute of Fundamental Research, \\ Shivakote, Hesaraghatta, Bangalore 560089, India.}
\emailAdd{kaustubh.singhi@icts.res.in}

\abstract
{
We study thermal two- and three-point functions of scalar primaries in holographic CFTs dual to AdS black branes.
The three-point function arises from a cubic scalar interaction in the bulk.
At vanishing external momenta, the tree-level results can be written in closed form and are exact in the operator dimensions.
We reproduce them through direct CFT calculations in two dimensions.
Their large-dimension expansions contain subleading exponential terms encoding the radial proper time from the horizon to the black hole singularity.
These results provide an exact test of earlier WKB and saddle-point predictions for higher-point thermal correlators, confirming the proper-time signature while correcting the predicted asymptotic form.
}

\begin{document}

\maketitle

\section{Introduction}
The AdS/CFT correspondence \cite{Maldacena:1997re,Gubser:1998bc,Witten:1998qj} relates thermal CFT correlators to bulk observables in black hole geometries.
The Euclidean setup contains only the exterior region, raising the question of how CFT observables encode the interior and its singularity.
This paper tests a specific prediction for higher-point thermal correlators:
whether their large-dimension behaviour contains the proper time from the horizon to the black hole singularity. \\

In \cite{Singhi:2024sdr}, I studied this question for scalar self-interactions and exchange interactions using WKB and saddle-point approximations.
The predicted leading exponential combined the geodesic distance between the boundary and the horizon with the proper time to the singularity.
Here, we test the contact three-point prediction at vanishing external momenta in the black brane geometry, where the bulk propagator and vertex integral can be treated exactly in the operator dimension.
The resulting large-dimension expansion encodes the time to singularity in
subleading exponential terms, while the leading exponential depends on the boundary-to-horizon distance.
This provides a direct benchmark for the earlier approximation in the zero-momentum case. \\

Our main results are the exact identical and mixed scalar contact three-point functions in general boundary dimension $d$, together with their analytically continued large-dimension limits.
We also rederive the thermal two-point function obtained by Berenstein and Mancilla \cite{Berenstein:2022nlj} and analyze its asymptotic behaviour.
We independently derive the $d = 2$ results by direct CFT calculation of the zero-momentum correlators. \\

The proper-time interpretation is motivated by exact one-point calculations.
Grinberg and Maldacena \cite{Grinberg:2020fdj} computed the thermal one-point function induced by coupling a bulk scalar to the square of the Weyl tensor.
Its analytically continued large-mass limit contains the proper time from the horizon to the singularity and agrees with a WKB and saddle-point calculation.
David and Kumar \cite{David:2022nfn} obtained further exact results for charged black branes in a large bulk spacetime dimension limit and for a hyperbolic black hole, with appropriate curvature couplings.\footnote{For a charged black hole with an inner horizon, the relevant geometric quantities are the proper time between the two horizons and the proper distance from the inner horizon to the singularity.}
Related signatures occur in curvature-induced thermal two-point contributions \cite{Rodriguez-Gomez:2021pfh,Krishna:2021fus}.
These examples motivate testing the higher-point approximation against the exact contact contributions considered here. \\

A complementary approach probes singularities by analytically continuing boundary correlators \cite{Kraus:2002iv,Fidkowski:2003nf,Festuccia:2005pi}.
For AdS--Schwarzschild black holes in four and higher dimensions, the geodesic approximation reveals a singularity on a secondary sheet of the two-sided correlator, associated with ``bouncing geodesics'': nearly null spacelike geodesics that connect the two boundaries and approach the black hole singularity.
Related signatures have been studied through the thermal OPE \cite{Ceplak:2024bja}, spectral functions \cite{Jia:2026ryl}, large-frequency two-point functions \cite{Afkhami-Jeddi:2025wra}, line defects \cite{Giombi:2026kdz}, and quasinormal modes \cite{Festuccia:2008zx,Arnaudo:2026tcy}.
Extensions to shock-wave geometries \cite{Horowitz:2023ury} and black holes with inner horizons \cite{Ceplak:2025dds,AliAhmad:2026wem} have also been studied.
For a charged black brane, exponentially small terms in the spectral density at large real frequency have been related to null geodesics reflecting off its timelike singularity \cite{Chakravarty:2026fdm}.
Analytically continued, smeared thermal four-point functions provide a related probe of local scattering behind the horizon \cite{Chakravarty:2025ncy}.
The fate of bouncing singularities has also been examined at finite coupling in SYK models \cite{Dodelson:2025jff,Buric:2026qsp}, with finite string length effects in the NS5-brane $SL(2,\mathbb{R})_{k}/U(1)$ worldsheet coset model \cite{Buric:2026sit}, and at finite $G_N$ \cite{Komargodski:2026oob,Batra:2026gbq,Liu:2026Smoothing}. \\

The remainder of the paper is organized as follows.
Section \ref{sec: preliminaries} fixes the conventions and collects the results.
The bulk calculation and large-dimension analysis are performed in section \ref{sec: gravity calculation}.
Section \ref{sec: 2d CFT} gives the CFT check in two dimensions.
We conclude with a short discussion of the comparison with the previous approximation and open questions in section \ref{sec: discussion}.

\section{Preliminaries and results}
\label{sec: preliminaries}
This section describes the setup, fixes notation, and collects the results obtained in this paper.
The physical observables of interest in this work are thermal correlation functions of primary operators in holographic CFTs.
Specifically, we consider a Euclidean CFT on $S_{\beta}^{1} \times \mathbb{R}^{d - 1}$, where $\beta$ is the inverse temperature.
The bulk dual is a black brane geometry with Hawking temperature $T = \frac{1}{\beta}$.
We use $\mathcal{O}$ for a scalar primary operator of conformal dimension $\Delta$.
Its dual is a scalar field $\phi$ propagating in the bulk spacetime, with mass $m$ related to the conformal dimension by \cite{Gubser:1998bc,Witten:1998qj}
\begin{gather}
    \label{eq: mass and conformal dimension relation}
    \Delta = \frac{d}{2} + \sqrt{\frac{d^{2}}{4} + m^{2}R^{2}},
\end{gather}
where $R$ is the AdS radius. \\

The translation symmetries of the background make it convenient to work with correlators in Fourier space.
The $n$-point thermal correlator of $\mathcal{O}$ is denoted by\footnote{We use the same symbol for operators in position and momentum space; the distinction is clear from the context.}
\begin{gather}
    \label{eq: momentum space correlator}
    \ev{\mathcal{O}(k_{1}) \cdots \mathcal{O}(k_{n})}_{\beta}.
\end{gather}
Here, $k_{i} \equiv (\omega_{i},\vec{k}_{i})$ are the momenta, including the Matsubara frequencies.
Translation symmetry enforces momentum conservation in the correlator \eqref{eq: momentum space correlator}.
Therefore, following the convention of \cite{Bzowski:2013sza}, we define a reduced correlator $\langle \! \ev{\cdot} \! \rangle$ via the relation
\begin{gather}
    \label{eq: reduced correlator}
    \ev{\mathcal{O}(k_{1}) \cdots \mathcal{O}(k_{n})}_{\beta} = (2\pi)^{d} \delta^{(d)}(k_{1} + \cdots + k_{n}) \langle \! \ev{\mathcal{O}(k_{1}) \cdots \mathcal{O}(k_{n})} \! \rangle_{\beta},
\end{gather}
where it is understood that time translation symmetry on $S_{\beta}^{1}$ requires the sum over Matsubara frequencies to vanish:
\begin{gather}
    2\pi\delta(\omega_{1} + \cdots + \omega_{n}) \to \beta\delta_{\omega_{1} + \cdots + \omega_{n},0}.
\end{gather}

We set all external momenta to zero and use the exact bulk-boundary propagator in this limit.
The thermal two-point function is given in \eqref{eq: two-point function gravity 2}; the identical and mixed contact three-point functions (see Figures \ref{fig: three-point Witten diagram} and \ref{fig: mixed Witten diagram}) are given in \eqref{eq: three-point function gravity 3} and \eqref{eq: mixed correlator gravity 4}.
Section \ref{sec: 2d CFT} checks these expressions using a direct CFT computation in two dimensions. \\

The correlators have poles for real positive values of $\Delta$.
So, to analyze the large conformal dimension behaviour, we give $\Delta$ a negative imaginary part and take the limit $\Delta \approx mR \to \infty (1 - i\epsilon)$.
The resulting expressions, given in \eqref{eq: two-point function expansion 1} and \eqref{eq: three-point function expansion 1}, are:
\begin{align}
    \label{eq: large conformal dimension behaviour 1}
    \langle \! \ev{\mathcal{O}(0) \mathcal{O}(0)} \! \rangle_{\beta} & \sim e^{-2m \ell_{h}} (1 - 2\mathfrak{q}^{2} + 2\mathfrak{q}^{4} - 2\mathfrak{q}^{6} + 2\mathfrak{q}^{8} + \cdots), \\
    \label{eq: large conformal dimension behaviour 2}
    \langle \! \ev{\mathcal{O}(0) \mathcal{O}(0) \mathcal{O}(0)} \! \rangle_{\beta} & \sim e^{-3m \ell_{h}} (1 - 3\mathfrak{q} + 3\mathfrak{q}^{2} - 3\mathfrak{q}^{4} + 3\mathfrak{q}^{5} + \cdots),
\end{align}
where $\mathfrak{q} = e^{-i\Delta\frac{\pi}{d}} \approx e^{-im\tau_{s}}$, with $\tau_{s} = \frac{\pi}{d}R$ being the time to singularity and $\ell_{h}$ the regularized distance between the asymptotic boundary and the black hole horizon; see equations \eqref{eq: bb regularized distance} and \eqref{eq: bb time to singularity}.
In this sense, the thermal correlators at vanishing momenta probe the bulk geometry, including the black hole interior and singularity.
The mixed thermal three-point function of two different primaries shows a similar signature when one of the operator dimensions is taken large; see \eqref{eq: mixed correlator gravity expansion 1}.

\section{Gravity calculation}
\label{sec: gravity calculation}

The bulk geometry dual to a holographic CFT on $S_{\beta}^{1} \times \mathbb{R}^{d - 1}$ is a Euclidean black brane whose metric is given by
\begin{gather}
    \label{eq: bb metric z variable}
    ds^{2} = \frac{R^{2}}{z^{2}} \left(f(z)d\tau^{2} + \frac{dz^{2}}{f(z)} + d\vec{x}^{\,2} \right), \quad f(z) = 1 - \frac{z^{d}}{z_{0}^{d}}.
\end{gather}
Here, $R$ is the AdS radius and $z_{0} = \frac{\beta d}{4\pi}$ is the location of the horizon.
The asymptotic boundary is at $z \to 0$.
We consider the following action for a bulk scalar field $\phi$, dual to a boundary primary operator $\mathcal{O}$:
\begin{gather}
    \label{eq: cubic scalar action}
    S[\phi] = \int{d^{d + 1}x}\sqrt{g} \left[ \frac{1}{2}\partial_{\mu}\phi \partial^{\mu}\phi + \frac{1}{2}m^{2}\phi^{2} + \frac{\lambda}{3!}\phi^{3} \right].
\end{gather}
The relation between the mass of the bulk scalar and the conformal dimension of the boundary operator was given in \eqref{eq: mass and conformal dimension relation}.
The self-interaction in \eqref{eq: cubic scalar action} provides a cubic vertex for Witten diagrams \cite{Gubser:1998bc,Witten:1998qj}. \\

Let us first focus on the free (quadratic) part of the action, which gives the Klein--Gordon equation for the massive scalar field.
In momentum space, with $k \equiv (\omega,\vec{k})$, it reduces to a radial equation:
\begin{gather}
    \label{eq: KG equation}
    z^{d + 1}\partial_{z}(z^{-d + 1} f(z) \partial_{z} \phi_{k}(z)) - \left( \frac{\omega^{2}}{f(z)}z^{2} + \vec{k}^{\,2}z^{2} + m^{2}R^{2} \right) \phi_{k}(z) = 0.
\end{gather}
The bulk-boundary propagator, $G_{k}(z)$, is defined as a regular solution to this equation with the boundary condition
\begin{gather}
    \label{eq: bulk-boundary prop bc}
    G_{k}(z) \overset{z \to 0}{\sim} z^{d - \Delta}.
\end{gather}
At vanishing momenta, the propagator is known in closed form as a hypergeometric function \cite{Grinberg:2020fdj}:
\begin{gather}
    \label{eq: bulk-boundary prop 2f1}
    G_{0}(z) \equiv G_{0}(w) = z_{0}^{d - \Delta} \frac{\Gamma(h)^{2}}{\Gamma(2h - 1)} w^{h} {}_{2}F_{1}(h,h ; 1 ; 1 - w),
\end{gather}
where $h = \frac{\Delta}{d}$ and we have defined a new radial variable
\begin{gather}
    \label{eq: defining w variable}
    w = \frac{z^{d}}{z_{0}^{d}}.
\end{gather}
An alternative form of the propagator uses the Legendre function \cite{Yarom:2009mw}:
\begin{gather}
    \label{eq: bulk-boundary prop leg}
    G_{0}(y) = z_{0}^{d - \Delta} \frac{\Gamma(h)^{2}}{\Gamma(2h - 1)} P_{h - 1}(y), \quad y = \frac{2}{w} - 1.
\end{gather}

\subsection{Two-point function}
\label{sec: gravity two-point}
Following the standard rules of AdS/CFT \cite{Freedman:1998tz,Son:2002sd} and holographic renormalization \cite{Skenderis:2002wp}, the thermal two-point function of the operator $\mathcal{O}$ can directly be extracted from the bulk-boundary propagator.
This computation appeared in \cite{Berenstein:2022nlj}.
The idea is to expand the propagator around $z = 0$ as
\begin{gather}
    \label{eq: expansion around z = 0}
    G_{k}(z) = z^{d - \Delta} (1 + \cdots) + B_{k} z^{\Delta} (1 + \cdots),
\end{gather}
where the $\cdots$ denote higher powers of $z$.
$B_{k}$ is termed the \textit{normalizable coefficient}.
The renormalized two-point function is then given by
\begin{gather}
    \label{eq: ren two-point function}
    \langle \! \ev{\mathcal{O}(k) \mathcal{O}(-k)} \! \rangle_{\beta} = (2\Delta - d) R^{d - 1} B_{k}.
\end{gather}
Using the connection formula for the hypergeometric function \cite[equation (15.8.4)]{NIST:DLMF}, we can extract the near-boundary behaviour of the bulk-boundary propagator at zero momentum:
\begin{gather}
    \label{eq: 2f1 expansion}
    w^{h} {}_{2}F_{1}(h,h ; 1 ; 1 - w) = \frac{\Gamma(2h - 1)}{\Gamma(h)^{2}}w^{1 - h}(1 + \cdots) + \frac{\Gamma(1 - 2h)}{\Gamma(1 - h)^{2}} w^{h}(1 + \cdots).
\end{gather}
Substituting this in \eqref{eq: bulk-boundary prop 2f1} and reading off the coefficient of $z^{\Delta}$ gives the normalizable coefficient $B_{0}$:
\begin{gather}
    \label{eq: normalizable coeff}
    B_{0} = z_{0}^{d - 2\Delta} \frac{\Gamma(h)^{2}\Gamma(1 - 2h)}{\Gamma(1 - h)^{2}\Gamma(2h - 1)}.
\end{gather}
We then have the two-point function at vanishing momenta
\begin{gather}
    \label{eq: two-point function gravity 1}
    \langle \! \ev{\mathcal{O}(0)\mathcal{O}(0)} \! \rangle_{\beta} = (2\Delta - d) z_{0}^{d - 2\Delta} R^{d - 1} \frac{\Gamma(h)^{2}\Gamma(1 - 2h)}{\Gamma(1 - h)^{2}\Gamma(2h - 1)}. 
\end{gather}
This reproduces the static response function obtained in equations (25)--(26) of \cite{Berenstein:2022nlj}.
Using the reflection formula, $\Gamma(z) \Gamma(1 - z) = \frac{\pi}{\sin \pi z}$, we write \eqref{eq: two-point function gravity 1} as
\begin{gather}
    \label{eq: two-point function gravity 2}
    \langle \! \ev{\mathcal{O}(0)\mathcal{O}(0)} \! \rangle_{\beta} = \frac{d z_{0}^{d - 2\Delta} R^{d - 1}}{\pi} \left( \frac{\Gamma(h)^{2}}{\Gamma(2h - 1)} \right)^{2} \frac{\sin^{2}\pi h}{\sin(2\pi h)}. 
\end{gather}

\paragraph{Large conformal dimension limit}
~\\
The thermal correlator \eqref{eq: two-point function gravity 2}, as a function of the conformal dimension, has poles at
\begin{gather}
    \label{eq: two-point function poles 1}
    \Delta = nd + \frac{d}{2}, \quad n = 1,2,3, \ldots.
\end{gather}
To avoid these poles in the asymptotic limit, we give a small negative imaginary part to $\Delta$ (equivalently $m$).
Since the spacetime dimension is fixed, this is implemented by $h \to \infty(1 - i\epsilon)$.
It is convenient to define the following variable to organize the expansion:
\begin{gather}
    \label{eq: defining q 1}
    \mathfrak{q} = e^{-i\pi h} = e^{-i\Delta \frac{\pi}{d}} \approx e^{-imR \frac{\pi}{d}}.
\end{gather}
Keeping terms exponential in $h$ and suppressing the temperature dependence, the asymptotic form of the two-point function is
\begin{gather}
    \label{eq: two-point function expansion 1}
    \langle \! \ev{\mathcal{O}(0)\mathcal{O}(0)} \! \rangle_{\beta} \sim 2^{-4h} \frac{(1 - \mathfrak{q}^{2})^{2}}{1 - \mathfrak{q}^{4}} = 2^{-4h} (1 - 2\mathfrak{q}^{2} + 2\mathfrak{q}^{4} - 2\mathfrak{q}^{6} + 2\mathfrak{q}^{8} + \cdots).
\end{gather}
We can identify the factors appearing in the expansion with geometric quantities in the black brane background.
The regularized distance between the asymptotic boundary and the horizon is
\begin{gather}
    \label{eq: bb regularized distance}
    \ell_{h} = \frac{R}{d} \lim_{w_{c} \to 0} \left( \int_{w_{c}}^{1}{\frac{dw}{w \sqrt{1 - w}}} + \log w_{c} \right) = \frac{R}{d} \log4.
\end{gather}
The proper time between the horizon and black hole singularity along a radially infalling geodesic is
\begin{gather}
    \label{eq: bb time to singularity}
    \tau_{s} = \frac{R}{d} \int_{1}^{\infty}{\frac{dw}{w \sqrt{w - 1}}} = \frac{\pi R}{d}.
\end{gather}
Thus, at large mass, the expansion \eqref{eq: two-point function expansion 1} can be expressed as
\begin{gather}
    \label{eq: two-point function expansion 2}
    \langle \! \ev{\mathcal{O}(0)\mathcal{O}(0)} \! \rangle_{\beta} \sim e^{-2m\ell_{h}} (1 - 2e^{-2im\tau_{s}} + 2e^{-4im\tau_{s}} + \cdots),
\end{gather}
where it is understood that the coefficient of each exponential factor inside the brackets is a power series in $m$.

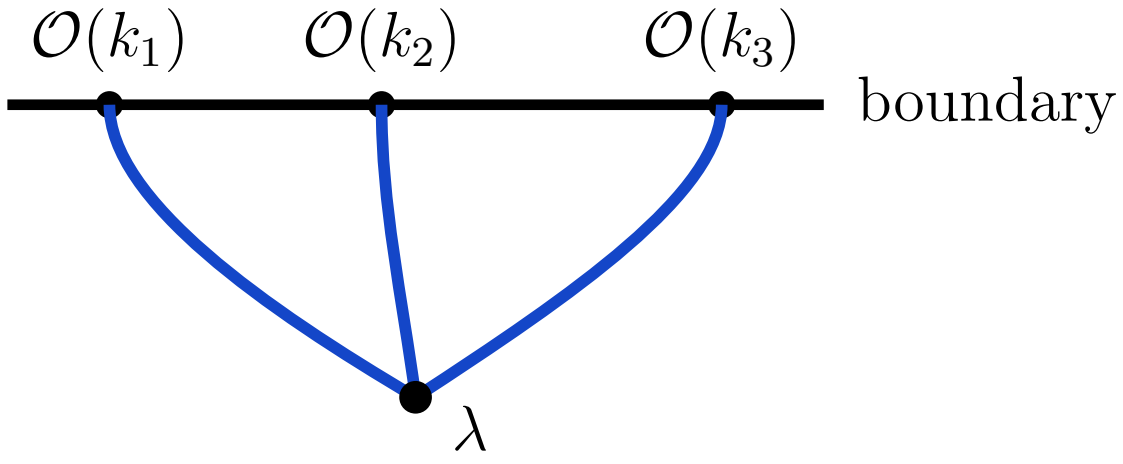
\begin{figure}[!t]
    \centering
    \begin{tikzpicture}[scale=0.45]
    \definecolor{bulkblue}{RGB}{20,70,200}
    
    \draw[line width=2pt] (-6,0) -- (6,0);
    
    \node at (8.4,0) {boundary};

    \coordinate (V) at (0,-4.3);
    
    \fill (-4.5,0) circle (0.2);
    \fill (-0.5,0) circle (0.2);
    \fill ( 4.5,0) circle (0.2);
    
    \node at (-4.5,0.95) {$\mathcal{O}(k_{1})$};
    
    \node at (-0.5,0.95) {$\mathcal{O}(k_{2})$};
    
    
    \node at (4.5,0.95) {$\mathcal{O}(k_{3})$};
    
    \draw[bulkblue,line width=2.2pt]
    (-4.5,0)
    .. controls (-4.5,-1.4) and (-2.2,-3.0)
    .. (V);
    
    \draw[bulkblue,line width=2.2pt]
    (-0.5,0)
    .. controls (-0.5,-1.5) and (-0.2,-2.8)
    .. (V);
    
    \draw[bulkblue,line width=2.2pt]
    (4.5,0)
    .. controls (4.5,-1.3) and (2.3,-2.8)
    .. (V);

    
    \fill[black] (V) circle (0.24);
    
    \node at (0.8,-4.75) {$\lambda$};
    
    \end{tikzpicture}
    
    \caption{The contact Witten diagram sourcing the three-point function of a scalar primary $\mathcal{O}$.
    The blue edges correspond to bulk-boundary propagators of the dual field $\phi$.
    The interaction vertex is integrated over the bulk Euclidean geometry.}
    \label{fig: three-point Witten diagram}
\end{figure}

\subsection{Three-point function}
\label{sec: gravity three-point}
The cubic self-interaction vertex in \eqref{eq: cubic scalar action} sources the three-point function of the operator $\mathcal{O}$.
The contact Witten diagram, shown in Figure \ref{fig: three-point Witten diagram}, evaluates to
\begin{gather}
    \label{eq: three-point function gravity 1}
    \langle \! \ev{\mathcal{O}(k_{1})\mathcal{O}(k_{2})\mathcal{O}(k_{3})} \! \rangle_{\beta} = - \lambda R^{d + 1} \int_{0}^{z_{0}}{\frac{dz}{z^{d + 1}}} G_{k_{1}}(z) G_{k_{2}}(z) G_{k_{3}}(z).
\end{gather}
Substituting \eqref{eq: bulk-boundary prop leg} for the zero-momentum propagator and changing the integration variable, we obtain the correlator at vanishing momenta:
\begin{gather}
    \label{eq: three-point function gravity 2}
    \langle \! \ev{\mathcal{O}(0)\mathcal{O}(0)\mathcal{O}(0)} \! \rangle_{\beta} = - \frac{\lambda z_{0}^{2d - 3\Delta} R^{d + 1}}{2d} \left( \frac{\Gamma(h)^{2}}{\Gamma(2h - 1)} \right)^{3} \int_{1}^{\infty}{dy} \, (P_{h - 1}(y))^{3}.
\end{gather}
The near-boundary behaviour of the integrand is $y^{3h - 3}$ as $y \to \infty$.
This means that the integral converges only if $3\Re(h) - 3 < -1 \Rightarrow \Re(h) < \frac{2}{3}$.
For larger values of $\Re(h)$, the integral is defined by analytic continuation in $h$. \\

Reference \cite{Bros:2009bz} computed the integral of three associated Legendre functions of the same order.\footnote{Interestingly, \cite{Bros:2009bz} encountered these integrals in the analysis of three-particle decay in de Sitter spacetimes.}
We adapt the general formula given in equations (1)--(2) of \cite{Bros:2009bz} to the integral of three Legendre functions:
\begin{gather}
    \label{eq: leg integral 1}
    \int_{1}^{\infty}{dy} \, P_{2a - \frac{1}{2}}(y) P_{2b - \frac{1}{2}}(y) P_{2c - \frac{1}{2}}(y) = \frac{1}{4\pi^{2}} \frac{\Gamma(\frac{1}{4} \pm a \pm b \pm c)}{\Gamma(\frac{1}{2} \pm 2a) \Gamma(\frac{1}{2} \pm 2b) \Gamma(\frac{1}{2} \pm 2c)},
\end{gather}
where \textit{each} $\pm$ in the arguments on the right-hand side denotes a product of Gamma functions over the different choices.
So, the numerator of \eqref{eq: leg integral 1} contains eight Gamma functions while the denominator contains six.
The integral in \eqref{eq: three-point function gravity 2} is obtained by setting $a = b = c = \frac{h}{2} - \frac{1}{4}$.
Using the reflection identity and the duplication formula, $\Gamma(z) \Gamma(z + \frac{1}{2}) = 2^{1 - 2z} \sqrt{\pi} \, \Gamma(2z)$, we can simplify the integral as follows:
\begin{gather}
    \label{eq: leg integral 2}
    \int_{1}^{\infty}{dy} \, (P_{h - 1}(y))^{3} = \frac{1}{\pi^{2}} \frac{\Gamma(3h - 1)\Gamma(\frac{h}{2})^{6}}{\Gamma(h)^{3} \Gamma(\frac{3h}{2})^{2}} \frac{\sin^{3}(\frac{\pi h}{2})}{\sin(\frac{3\pi h}{2})}.
\end{gather}
Substituting \eqref{eq: leg integral 2} in \eqref{eq: three-point function gravity 2} gives the thermal three-point function:
\begin{gather}
    \label{eq: three-point function gravity 3}
    \langle \! \ev{\mathcal{O}(0)\mathcal{O}(0)\mathcal{O}(0)} \! \rangle_{\beta} = - \frac{\lambda z_{0}^{2d - 3\Delta} R^{d + 1}}{2\pi^{2}d} \frac{\Gamma(3h - 1) \Gamma(h)^{3} \Gamma(\frac{h}{2})^{6}}{\Gamma(2h - 1)^{3} \Gamma(\frac{3h}{2})^{2}} \frac{\sin^{3}(\frac{\pi h}{2})}{\sin(\frac{3\pi h}{2})}.
\end{gather}

\paragraph{Large conformal dimension limit}
~\\
As with the two-point function, we now study the large conformal dimension behaviour of the thermal three-point function \eqref{eq: three-point function gravity 3}.
From the ratio of sine functions, we identify the poles of the correlator at positive real values of the conformal dimension:\footnote{The three-point function has a pole at $\Delta = \frac{d}{3}$ as well, coming from the factor $\Gamma(3h - 1)$ in \eqref{eq: three-point function gravity 3}, but this is below the Breitenlohner--Freedman (BF) bound \cite{Breitenlohner:1982bm} $\Delta \geq \frac{d}{2}$.}
\begin{gather}
    \label{eq: three-point function poles 1}
    \Delta = \frac{2nd}{3},
\end{gather}
where $n \geq 1$ is an integer not divisible by 3.
Again, to avoid these poles, we analyze the limit $h \to \infty(1 - i\epsilon)$ of the correlator.
The terms exponential in the conformal dimension come from the ratios of Gamma functions and sine functions.
In the variable $\mathfrak{q}$ defined in \eqref{eq: defining q 1}, the thermal three-point function is expanded as
\begin{gather}
    \label{eq: three-point function expansion 1}
    \langle \! \ev{\mathcal{O}(0)\mathcal{O}(0)\mathcal{O}(0)} \! \rangle_{\beta} \sim 2^{-6h} \frac{(1 - \mathfrak{q})^{3}}{1 - \mathfrak{q}^{3}} = 2^{-6h} (1 - 3\mathfrak{q} + 3\mathfrak{q}^{2} - 3\mathfrak{q}^{4} + 3\mathfrak{q}^{5} + \cdots).
\end{gather}
The expansion can be re-expressed using the geometric quantities \eqref{eq: bb regularized distance} and \eqref{eq: bb time to singularity} as
\begin{gather}
    \label{eq: three-point function expansion 2}
    \hspace{-0.35cm}\langle \! \ev{\mathcal{O}(0)\mathcal{O}(0)\mathcal{O}(0)} \! \rangle_{\beta} \sim e^{-3m\ell_{h}} (1 - 3e^{-im\tau_{s}} + 3e^{-2im\tau_{s}} - 3e^{-4im\tau_{s}} + 3e^{-5im\tau_{s}} + \cdots).
\end{gather}
This indicates that higher-point thermal correlators contain an imprint of the black hole singularity.
However, the result \eqref{eq: three-point function expansion 2} differs from the form $e^{-3m(\ell_{h} + i\tau_{s})}$ suggested in \cite{Singhi:2024sdr}.
In fact, the expansion \eqref{eq: three-point function expansion 2} has no terms corresponding to $e^{-3lim\tau_{s}}$ for any positive integer $l$.
See section \ref{sec: discussion} for further discussion.

\begin{figure}[!t]
    \centering
    \begin{tikzpicture}[scale=0.45]
    \definecolor{bulkblue}{RGB}{20,70,200}
    \definecolor{bulkred}{RGB}{180,20,20}
    
    \draw[line width=2pt] (-6,0) -- (6,0);
    
    \node at (8.4,0) {boundary};

    \coordinate (V) at (0,-4.3);
    
    \draw[bulkblue,line width=2.2pt]
    (-4.5,0)
    .. controls (-4.5,-1.4) and (-2.2,-3.0)
    .. (V);
    
    \draw[bulkblue,line width=2.2pt]
    (-0.5,0)
    .. controls (-0.5,-1.5) and (-0.2,-2.8)
    .. (V);
    
    \draw[bulkred,line width=2.2pt]
    (4.5,0)
    .. controls (4.5,-1.3) and (2.3,-2.8)
    .. (V);
    
    \fill (-4.5,0) circle (0.2);
    \fill (-0.5,0) circle (0.2);
    \fill ( 4.5,0) circle (0.2);
    
    \node at (-4.5,0.95) {$\mathcal{O}_{1}(k_{1})$};
    
    \node at (-0.5,0.95) {$\mathcal{O}_{1}(k_{2})$};
    
    
    \node at (4.5,0.95) {$\mathcal{O}_{2}(k_{3})$};

    \fill[black] (V) circle (0.24);
    
    \node at (0.8,-4.75) {$\kappa$};
    
    \end{tikzpicture}

    \caption{The contact Witten diagram sourcing the mixed three-point function of two primaries $\mathcal{O}_{1}$ and $\mathcal{O}_{2}$.
    The edges correspond to bulk-boundary propagators of the dual fields: blue for $\phi_{1}$ and red for $\phi_{2}$.
    The interaction vertex is integrated over the bulk Euclidean geometry.}
    \label{fig: mixed Witten diagram}
\end{figure}
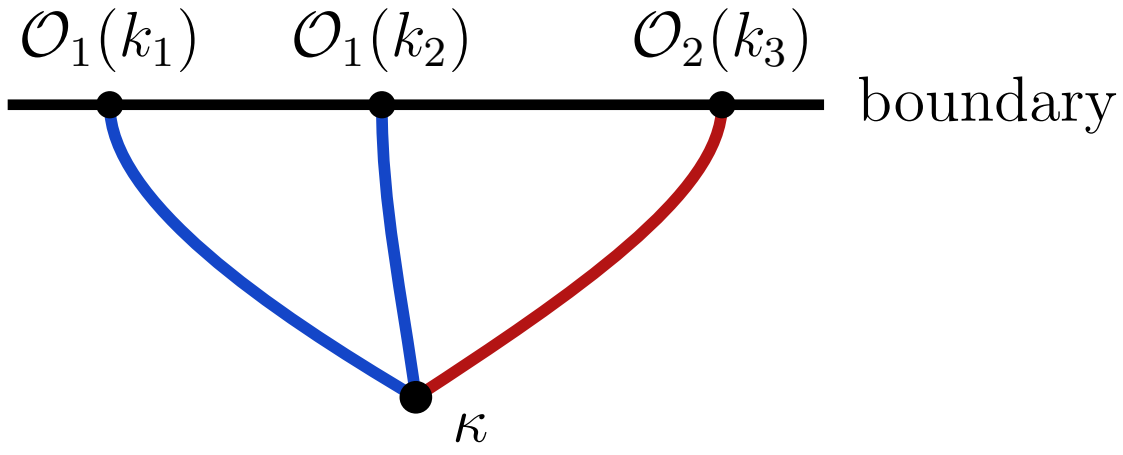

\subsection{Mixed correlator}
\label{sec: gravity mixed correlator}
In \cite{Singhi:2024sdr}, it was argued that factors of time to singularity may also appear in mixed thermal correlators of heavy and light operators.
We can test this by directly calculating the following three-point function at vanishing momenta:
\begin{gather}
    \label{eq: mixed correlator gravity 1}
    \langle \! \ev{\mathcal{O}_{1}(0) \mathcal{O}_{1}(0) \mathcal{O}_{2}(0)} \! \rangle_{\beta},
\end{gather}
where $\mathcal{O}_{1}$ and $\mathcal{O}_{2}$ are two scalar primaries holographically dual to scalar fields $\phi_{1}$ and $\phi_{2}$, respectively.
The conformal dimensions, $\Delta_{i} \equiv dh_{i}$, are related to the masses, $m_{i}$, of the dual scalars by the usual relation \eqref{eq: mass and conformal dimension relation}.
We consider the bulk action
\begin{gather}
    \label{eq: mixed bulk action}
    S[\phi_{1},\phi_{2}] = \frac{1}{2}\int{d^{d + 1}x}\sqrt{g}\left[ \partial_{\mu}\phi_{1}\partial^{\mu}\phi_{1} + \partial_{\mu}\phi_{2}\partial^{\mu}\phi_{2} + m_{1}^{2}\phi_{1}^{2} + m_{2}^{2}\phi_{2}^{2} + \kappa\phi_{1}^{2}\phi_{2} \right].
\end{gather}
The interaction $\phi_{1}^{2}\phi_{2}$ provides a cubic vertex for Witten diagrams.
The contact diagram using this vertex sources the three-point function $\ev{\mathcal{O}_{1} \mathcal{O}_{1} \mathcal{O}_{2}}$, as shown in Figure \ref{fig: mixed Witten diagram}.
It evaluates to the bulk integral
\begin{gather}
    \label{eq: mixed correlator gravity 2}
    \langle \! \ev{\mathcal{O}_{1}(k_{1}) \mathcal{O}_{1}(k_{2}) \mathcal{O}_{2}(k_{3})} \! \rangle_{\beta} = -\kappa R^{d + 1} \int_{0}^{z_{0}}{\frac{dz}{z^{d + 1}}} G_{1,k_{1}}(z) G_{1,k_{2}}(z) G_{2,k_{3}}(z),
\end{gather}
where we have denoted the bulk-boundary propagator of $\phi_{i}$ by $G_{i,k}(z)$.
Using the zero-momentum propagators \eqref{eq: bulk-boundary prop leg} and changing the integration variable, we express the correlator \eqref{eq: mixed correlator gravity 1} as
\begin{align}
    \begin{split}
        \label{eq: mixed correlator gravity 3}
        \langle \! \ev{\mathcal{O}_{1}(0) \mathcal{O}_{1}(0) \mathcal{O}_{2}(0)} \! \rangle_{\beta} & = -\frac{\kappa z_{0}^{2d - 2\Delta_{1} - \Delta_{2}} R^{d + 1}}{2d} \frac{\Gamma(h_{1})^{4} \Gamma(h_{2})^{2}}{\Gamma(2h_{1} - 1)^{2} \Gamma(2h_{2} - 1)} \\
        & \qquad \qquad \qquad \times \int_{1}^{\infty}{dy} \, P_{h_{1} - 1}(y) P_{h_{1} - 1}(y) P_{h_{2} - 1}(y).
    \end{split}
\end{align}
Near-boundary behaviour implies that the integral converges for $\Re(2h_{1} + h_{2}) < 2$.
For other values, we define it by analytic continuation in $h_{1}$ and $h_{2}$.

The general formula \eqref{eq: leg integral 1} can be used for \eqref{eq: mixed correlator gravity 3} by setting $a = b = \frac{h_{1}}{2} - \frac{1}{4}$ and $c = \frac{h_{2}}{2} - \frac{1}{4}$.
The resulting ratio of Gamma functions can be simplified as follows:
\begin{gather}
    \label{eq: leg integral 3}
    \int_{1}^{\infty}{dy} \, P_{h_{1} - 1}(y) P_{h_{1} - 1}(y) P_{h_{2} - 1}(y) = \frac{1}{8\pi^{2}} \frac{\Gamma(2h_{1} + h_{2} - 1) \Gamma(h_{1} - \frac{h_{2}}{2})^{2} \Gamma(\frac{h_{2}}{2})^{4}}{\Gamma(h_{1} + \frac{h_{2}}{2})^{2} \Gamma(2h_{1} - h_{2}) \Gamma(h_{2})^{2}} \, \mathcal{S},
\end{gather}
where we have defined
\begin{gather}
    \label{eq: leg integral 4}
    \mathcal{S} = \frac{\sin^{2}\pi h_{1} \sin\pi h_{2}}{\sin(\frac{\pi(2h_{1} + h_{2})}{2}) \cos(\frac{\pi(2h_{1} - h_{2})}{2}) \cos^{2}(\frac{\pi h_{2}}{2})}.
\end{gather}
Substituting \eqref{eq: leg integral 3} in \eqref{eq: mixed correlator gravity 3} gives the thermal mixed correlator:
\begin{gather}
    \label{eq: mixed correlator gravity 4}
    \langle \! \ev{\mathcal{O}_{1}(0) \mathcal{O}_{1}(0) \mathcal{O}_{2}(0)} \! \rangle_{\beta} = -\frac{\kappa z_{0}^{2d - 2\Delta_{1} - \Delta_{2}} R^{d + 1}}{16\pi^{2}d} \, \mathcal{A},
\end{gather}
where we have introduced
\begin{gather}
    \label{eq: mixed correlator gravity 5}
    \mathcal{A} = \frac{\Gamma(2h_{1} + h_{2} - 1) \Gamma(h_{1} - \frac{h_{2}}{2})^{2} \Gamma(\frac{h_{2}}{2})^{4} \Gamma(h_{1})^{4}}{\Gamma(h_{1} + \frac{h_{2}}{2})^{2} \Gamma(2h_{1} - h_{2}) \Gamma(2h_{1} - 1)^{2} \Gamma(2h_{2} - 1)} \, \mathcal{S}.
\end{gather}

\paragraph{Large conformal dimension limit}
~\\
We wish to analyze the behaviour of the mixed correlator at large $\Delta_{1}$ while holding $\Delta_{2}$ fixed.
For generic values of $\Delta_{2}$, the correlator has poles at
\begin{gather}
    \label{eq: mixed correlator gravity poles}
    \Delta_{1} = nd - \frac{\Delta_{2}}{2} \quad \text{and} \quad 
    \Delta_{1} = \left( n + \frac{1}{2} \right)d + \frac{\Delta_{2}}{2},
\end{gather}
where $n$ is an integer such that the BF bound, $\Delta_{1} \geq \frac{d}{2}$, is satisfied.
As before, we avoid the poles by taking the limit along a ray in the lower half of the complex plane, $\Delta_{1} \to \infty(1 - i\epsilon)$.
The expansion of the mixed correlator begins as
\begin{gather}
    \label{eq: mixed correlator gravity expansion 1}
    \langle \! \ev{\mathcal{O}_{1}(0) \mathcal{O}_{1}(0) \mathcal{O}_{2}(0)} \! \rangle_{\beta} \sim 2^{-4h_{1}} (1 - \mathfrak{q}_{1}^{2}(2 + \mathfrak{q}_{2}^{-1} - \mathfrak{q}_{2}) + O(\mathfrak{q}_{1}^{4})),
\end{gather}
where we have defined
\begin{gather}
    \label{eq: defining q 2}
    \mathfrak{q}_{i} = e^{-i\pi h_{i}} = e^{-i\Delta_{i} \frac{\pi}{d}}.
\end{gather}
Identifying the geometric quantities \eqref{eq: bb regularized distance} and \eqref{eq: bb time to singularity}, we write
\begin{gather}
    \label{eq: mixed correlator gravity expansion 2}
    \langle \! \ev{\mathcal{O}_{1}(0) \mathcal{O}_{1}(0) \mathcal{O}_{2}(0)} \! \rangle_{\beta} \sim e^{-2m_{1}\ell_{h}} (1 - e^{-2im_{1}\tau_{s}}(2 + \mathfrak{q}_{2}^{-1} - \mathfrak{q}_{2}) + O(e^{-4im_{1}\tau_{s}})).
\end{gather}
We see that the thermal mixed correlator also contains a signature of the black hole singularity.
The precise form \eqref{eq: mixed correlator gravity expansion 2} differs from the one suggested in \cite{Singhi:2024sdr}, where the expected form was $e^{-2m_{1}(\ell_{h} + i\tau_{s})}$.

\section{$d = 2$ CFT}
\label{sec: 2d CFT}
In $d = 2$, thermal correlators on $S_{\beta}^{1} \times \mathbb{R}$ follow from the vacuum ones on the plane by a conformal map; see equation \eqref{eq: conformal map}.
The functional form of the two- and three-point correlators on the plane can be fixed completely using conformal symmetry.
Moreover, the remaining data --- normalization and three-point OPE coefficients --- can be computed for a holographic theory using bulk Witten diagrams in the vacuum AdS geometry.
Momentum-space thermal three-point functions in two dimensions were previously computed in \cite{Becker:2014jla} with Lorentzian ordering prescriptions.
Here, we evaluate the zero-momentum Euclidean correlators, providing a separate check of the bulk formulas. \\

Let $\xi = \sigma + i\tau, \bar{\xi} = \sigma - i\tau$ be coordinates on the cylinder $S_{\beta}^{1} \times \mathbb{R}$, with $\sigma \in \mathbb{R}$ and $\tau \sim \tau + \beta$.
The conformal map
\begin{gather}
    \label{eq: conformal map}
    z = \exp(\frac{2\pi}{\beta} \xi), \quad \bar{z} = \exp(\frac{2\pi}{\beta} \bar{\xi})
\end{gather}
takes the cylinder to the complex plane $\mathbb{C}\backslash\{0\}$.\footnote{Here, the $\tau$-variable on the cylinder is Euclidean time, so the correlation functions are thermal correlators.
If instead $\tau$ is regarded as a compact spatial direction, then the same correlation functions are vacuum correlators on a spatial circle.} A primary $\mathcal{O}$ with conformal weights $(h,\bar{h})$ transforms under the map \eqref{eq: conformal map} as:
\begin{gather}
    \label{eq: primary transformation}
    \mathcal{O}(\xi(z),\bar{\xi}(\bar{z})) = \left( \frac{\partial z}{\partial \xi} \right)^{h} \left( \frac{\partial \bar{z}}{\partial \bar{\xi}} \right)^{\bar{h}} \mathcal{O}(z,\bar{z}) = \left( \frac{2\pi}{\beta} \right)^{h + \bar{h}} z^{h} \bar{z}^{\bar{h}} \mathcal{O}(z,\bar{z}).
\end{gather}
We now restrict to primaries with equal holomorphic and antiholomorphic weights: $h = \bar{h} = \frac{\Delta}{2}$, where, for a holographic theory, $\Delta$ is given by \eqref{eq: mass and conformal dimension relation}.
Further, we will use the compact notation $\mathcal{O}(z) \equiv \mathcal{O}(z,\bar{z})$ and $\mathcal{O}(\xi) \equiv \mathcal{O}(\xi,\bar{\xi})$ for operators on the plane and cylinder, respectively.

\subsection{Two-point function}
\label{sec: 2d two-point}
We start with the vacuum two-point function of $\mathcal{O}$ on the plane, which, by conformal invariance, is given by
\begin{gather}
    \label{eq: vacuum two-point function 1}
    \ev{\mathcal{O}(z_{1}) \mathcal{O}(z_{2})} = \frac{\mathcal{C}_{\mathcal{O}}}{\abs{z_{12}}^{4h}},
\end{gather}
where $z_{12} \equiv z_{1} - z_{2}$.
For a holographic theory, the normalization $\mathcal{C}_{\mathcal{O}}$ of the two-point function can be computed using the bulk-boundary propagator in vacuum AdS \cite[equation (96)]{Freedman:1998tz}:
\begin{gather}
    \label{eq: two-point function normalization}
    \mathcal{C}_{\mathcal{O}} = \frac{2R}{\pi} (2h - 1)^{2}.
\end{gather}
Applying the conformal transformation \eqref{eq: primary transformation} to \eqref{eq: vacuum two-point function 1} gives the thermal two-point function in position space:
\begin{gather}
    \label{eq: thermal two-point function 1}
    \ev{\mathcal{O}(\xi_{1}) \mathcal{O}(\xi_{2})}_{\beta} = \mathcal{C}_{\mathcal{O}} \left( \frac{2\pi}{\beta} \right)^{2\Delta} \frac{\abs{z_{1}}^{2h} \abs{z_{2}}^{2h}}{\abs{z_{12}}^{4h}}.
\end{gather}
While this expression can be further simplified to \cite{Son:2002sd}
\begin{gather}
    \label{eq: thermal two-point function 2}
    \ev{\mathcal{O}(\xi_{1}) \mathcal{O}(\xi_{2})}_{\beta} = \frac{\mathcal{C}_{\mathcal{O}} \left( \frac{2\pi}{\beta} \right)^{2\Delta}}{\abs{2\sinh(\frac{\pi}{\beta} \xi_{12})}^{4h}},
\end{gather}
with $\xi_{12} = \xi_{1} - \xi_{2}$, we will find it more convenient to work with \eqref{eq: thermal two-point function 1}. \\

The translation symmetry on the cylinder allows us to define the reduced Fourier space thermal two-point function
\begin{gather}
    \label{eq: thermal two-point fourier 1}
    \langle \! \ev{\mathcal{O}(\omega,k) \mathcal{O}(-\omega,-k)} \! \rangle_{\beta} = \int{d^{2}\xi} \, e^{-i(\omega\tau + k\sigma)} \ev{\mathcal{O}(\xi) \mathcal{O}(0)}_{\beta},
\end{gather}
where $d^{2}\xi \equiv d\sigma d\tau$.
Recall that we are interested in thermal correlators at vanishing momenta.
Substituting \eqref{eq: thermal two-point function 1} into \eqref{eq: thermal two-point fourier 1} and using the change of variables \eqref{eq: conformal map} from $\xi$ to $z$ gives
\begin{gather}
    \label{eq: thermal two-point fourier 2}
    \langle \! \ev{\mathcal{O}(0) \mathcal{O}(0)} \! \rangle_{\beta} = \mathcal{C}_{\mathcal{O}} \left( \frac{2\pi}{\beta} \right)^{2\Delta - 2} \int{d^{2}z} \abs{z}^{2h - 2} \abs{1 - z}^{-4h},
\end{gather}
where $d^{2}z \equiv d\Re(z) d\Im(z)$.
The integral in \eqref{eq: thermal two-point fourier 2} is an instance of the complex Selberg integral \cite{Aomoto:1987}, also sometimes referred to as the Dotsenko--Fateev integral \cite{Dotsenko:1984nm}:
\begin{gather}
    \label{eq: complex Selberg integral 1}
    S_{n}^{\mathbb{C}}(\alpha,\beta;\gamma) = \int{\prod_{k = 1}^{n} d^{2}z_{k}} \, \abs{z_{k}}^{2\alpha - 2} \abs{1 - z_{k}}^{2\beta - 2} \prod_{1 \leq k < l \leq n} \abs{z_{k} - z_{l}}^{4\gamma}.
\end{gather}
The general formula for $S_{n}^{\mathbb{C}}(\alpha,\beta;\gamma)$ can be written as \cite{Forrester:2007,Neretin:2022pbv}
\begin{align}
    \begin{split}
        \label{eq: complex Selberg integral 2}
        S_{n}^{\mathbb{C}}(\alpha,\beta;\gamma) & = n! \prod_{j = 0}^{n - 1} \bigg[ \frac{\sin(\pi(\alpha + j\gamma))\sin(\pi(\beta + j\gamma))\sin(\pi(j + 1)\gamma)}{\sin(\pi(\alpha+ \beta + (n + j - 1)\gamma)) \sin(\pi\gamma)} \\
        & \hspace{1.5cm} \left( \frac{\Gamma(\alpha + j\gamma) \Gamma(\beta + j\gamma) \Gamma((j + 1)\gamma)}{\Gamma(\alpha + \beta + (n + j - 1)\gamma) \Gamma(\gamma)}\right)^{2} \bigg].
    \end{split}
\end{align}
The defining integral \eqref{eq: complex Selberg integral 1} is convergent in a ``narrow domain'' \cite{Forrester:2007,Neretin:2022pbv}, but the result \eqref{eq: complex Selberg integral 2} in terms of the Gamma and trigonometric functions can be analytically continued to all complex parameter values. \\

The complex Selberg integral relevant for the thermal two-point function \eqref{eq: thermal two-point fourier 2} is $S_{1}^{\mathbb{C}}(h,1 - 2h;\gamma)$.\footnote{For $n = 1$, the complex Selberg integral does not depend on $\gamma$.} Using the general formula \eqref{eq: complex Selberg integral 2}, its value is given by
\begin{gather}
    \label{eq: two-point Selberg}
    S_{1}^{\mathbb{C}}(h,1 - 2h;\gamma) 
    = \left( \frac{\Gamma(h)^{2}}{\Gamma(2h)} \right)^{2} \frac{\sin^{2}\pi h}{\sin(2\pi h)},
\end{gather}
where we have used the reflection formula to simplify the expression.
Substituting the formula back into \eqref{eq: thermal two-point fourier 2} and using the holographic value \eqref{eq: two-point function normalization} for the normalization gives
\begin{gather}
    \label{eq: thermal two-point fourier 3}
    \langle \! \ev{\mathcal{O}(0) \mathcal{O}(0)} \! \rangle_{\beta} = \frac{2R}{\pi} \left( \frac{2\pi}{\beta} \right)^{2\Delta - 2} \left( \frac{\Gamma(h)^{2}}{\Gamma(2h - 1)} \right)^{2} \frac{\sin^{2}\pi h}{\sin(2\pi h)}.
\end{gather}
This precisely matches the gravity result \eqref{eq: two-point function gravity 2} for $d = 2$, in which case the black brane radius is related to the inverse temperature by $z_{0} = \frac{\beta}{2\pi}$.

\subsection{Three-point function}
\label{sec: 2d three-point}
We now turn to the three-point function of the primary $\mathcal{O}$.
Again, by conformal invariance, the vacuum three-point correlator on the plane takes the form
\begin{gather}
    \label{eq: vacuum three-point function 1}
    \ev{\mathcal{O}(z_{1}) \mathcal{O}(z_{2}) \mathcal{O}(z_{3})} = \frac{\mathcal{C}_{\mathcal{O}\mathcal{O}\mathcal{O}}}{\abs{z_{12}}^{2h} \abs{z_{23}}^{2h} \abs{z_{13}}^{2h}},
\end{gather}
where $z_{ij} = z_{i} - z_{j}$.
The factor $\mathcal{C}_{\mathcal{O}\mathcal{O}\mathcal{O}}$ is related to the OPE coefficient of the operator $\mathcal{O}$.
For a holographic theory, it can be computed in the semiclassical limit by bulk perturbation theory in the vacuum AdS geometry.
The vacuum contact Witten diagram (with the same schematic form as Figure \ref{fig: three-point Witten diagram}) corresponding to a cubic interaction of bulk scalars was computed in \cite{Freedman:1998tz}.
Specializing their formula \cite[equation (25)]{Freedman:1998tz} to the self-interacting scalar in $d = 2$ gives
\begin{gather}
    \label{eq: three-point coefficient}
    \mathcal{C}_{\mathcal{O}\mathcal{O}\mathcal{O}} = - \frac{\lambda R^{3}}{2\pi^{2}} \frac{\Gamma(3h - 1) \Gamma(h)^{3}}{\Gamma(2h - 1)^{3}}.
\end{gather}

We apply the conformal transformation \eqref{eq: primary transformation} to \eqref{eq: vacuum three-point function 1} to obtain the thermal three-point function in position space:
\begin{gather}
    \label{eq: thermal three-point function 1}
    \ev{\mathcal{O}(\xi_{1}) \mathcal{O}(\xi_{2}) \mathcal{O}(\xi_{3})}_{\beta} = \mathcal{C}_{\mathcal{O}\mathcal{O}\mathcal{O}} \left( \frac{2\pi}{\beta} \right)^{3\Delta} \frac{\abs{z_{1}}^{2h} \abs{z_{2}}^{2h} \abs{z_{3}}^{2h}}{\abs{z_{12}}^{2h} \abs{z_{23}}^{2h} \abs{z_{13}}^{2h}}.
\end{gather}
This formula can be expressed as
\begin{gather}
    \label{eq: thermal three-point function 2}
    \ev{\mathcal{O}(\xi_{1}) \mathcal{O}(\xi_{2}) \mathcal{O}(\xi_{3})}_{\beta} = \frac{\mathcal{C}_{\mathcal{O}\mathcal{O}\mathcal{O}} \left( \frac{2\pi}{\beta} \right)^{3\Delta}}{\abs{2\sinh(\frac{\pi}{\beta} \xi_{12})}^{2h} \abs{2\sinh(\frac{\pi}{\beta} \xi_{23})}^{2h} \abs{2\sinh(\frac{\pi}{\beta} \xi_{13})}^{2h}},
\end{gather}
with $\xi_{ij} = \xi_{i} - \xi_{j}$, but it is easier to work with \eqref{eq: thermal three-point function 1} when taking the Fourier transform. \\

The reduced Fourier space thermal three-point function is defined as
\begin{align}
    \begin{split}
        \label{eq: thermal three-point fourier 1}
        \langle \! \ev{\mathcal{O}(\omega_{1},k_{1}) \mathcal{O}(\omega_{2},k_{2}) \mathcal{O}(\omega_{3},k_{3})} \! \rangle_{\beta} & \\
        & \hspace{-2cm} = \int{d^{2}\xi_{1} d^{2}\xi_{2}} \, e^{-i(\omega_{1}\tau_{1} + k_{1}\sigma_{1} + \omega_{2}\tau_{2} + k_{2}\sigma_{2})} \ev{\mathcal{O}(\xi_{1}) \mathcal{O}(\xi_{2}) \mathcal{O}(0)}_{\beta}.
    \end{split}
\end{align}
For vanishing momenta, we get
\begin{align}
    \label{eq: thermal three-point fourier 2}
    \langle \! \ev{\mathcal{O}(0) \mathcal{O}(0) \mathcal{O}(0)} \! \rangle_{\beta} & = \mathcal{C}_{\mathcal{O}\mathcal{O}\mathcal{O}} \left( \frac{2\pi}{\beta} \right)^{3\Delta - 4} \\
    \notag & \quad \times \int{d^{2}z_{1} d^{2}z_{2}} \, \abs{z_{1}}^{2h - 2} \abs{z_{2}}^{2h - 2} \abs{1 - z_{1}}^{-2h} \abs{1 - z_{2}}^{-2h} \abs{z_{1} - z_{2}}^{-2h}.
\end{align}
The above integral is the complex Selberg integral $S_{2}^{\mathbb{C}}(h,1 - h;-h/2)$.
Specializing the general formula \eqref{eq: complex Selberg integral 2} to this case and using the reflection identity for the Gamma function, we obtain
\begin{gather}
    \label{eq: three-point Selberg}
    S_{2}^{\mathbb{C}}(h,1 - h;-h/2) = \frac{1}{2} \frac{\Gamma(\frac{h}{2})^{6}}{\Gamma(\frac{3h}{2})^{2}} \frac{\sin^{3}(\frac{\pi h}{2})}{\sin(\frac{3\pi h}{2})}.
\end{gather}
Substituting the formula back into \eqref{eq: thermal three-point fourier 2} and using the holographic value \eqref{eq: three-point coefficient} for $\mathcal{C}_{\mathcal{O}\mathcal{O}\mathcal{O}}$ gives the thermal three-point function
\begin{gather}
    \label{eq: thermal three-point fourier 3}
    \langle \! \ev{\mathcal{O}(0) \mathcal{O}(0) \mathcal{O}(0)} \! \rangle_{\beta} = - \frac{\lambda R^{3}}{4\pi^{2}} \left( \frac{2\pi}{\beta} \right)^{3\Delta - 4} \frac{\Gamma(3h - 1) \Gamma(h)^{3} \Gamma(\frac{h}{2})^{6}}{\Gamma(2h - 1)^{3} \Gamma(\frac{3h}{2})^{2}} \frac{\sin^{3}(\frac{\pi h}{2})}{\sin(\frac{3\pi h}{2})}.  
\end{gather}
This is in precise agreement with the gravity result \eqref{eq: three-point function gravity 3} for $d = 2$ and $z_{0} = \frac{\beta}{2\pi}$.

\subsection{Mixed correlator}
\label{sec: 2d mixed correlator}
We now repeat the CFT calculation for the mixed three-point function \eqref{eq: mixed correlator gravity 1}.
Conformal invariance fixes the vacuum form to
\begin{gather}
    \label{eq: mixed vacuum three-point function}
    \ev{\mathcal{O}_{1}(z_{1}) \mathcal{O}_{1}(z_{2}) \mathcal{O}_{2}(z_{3})}
    = \frac{\mathcal{C}_{112}}{\abs{z_{12}}^{4h_{1} - 2h_{2}} \abs{z_{13}}^{2h_{2}} \abs{z_{23}}^{2h_{2}}},
\end{gather}
where the vacuum contact diagram for the interaction in \eqref{eq: mixed bulk action} gives \cite{Freedman:1998tz}
\begin{gather}
    \label{eq: mixed vacuum three-point coefficient}
    \mathcal{C}_{112} = -\frac{\kappa R^{3}}{2\pi^{2}}
    \frac{\Gamma(2h_{1} + h_{2} - 1)\Gamma(2h_{1} - h_{2})\Gamma(h_{2})^{2}}
    {\Gamma(2h_{1} - 1)^{2}\Gamma(2h_{2} - 1)}.
\end{gather}
Applying \eqref{eq: primary transformation} gives the thermal correlator.
Its zero-momentum Fourier transform, as before, is given by a complex Selberg integral:
\begin{align}
    \label{eq: mixed thermal three-point fourier}
    \langle \! \ev{\mathcal{O}_{1}(0) \mathcal{O}_{1}(0) \mathcal{O}_{2}(0)} \! \rangle_{\beta} = \mathcal{C}_{112}\left(\frac{2\pi}{\beta}\right)^{2\Delta_{1} + \Delta_{2} - 4} S_{2}^{\mathbb{C}}(h_{1},1 - h_{2};h_{2}/2 - h_{1}).
\end{align}
Using \eqref{eq: complex Selberg integral 2} and the reflection formula gives
\begin{gather}
    \label{eq: mixed three-point Selberg}
    S_{2}^{\mathbb{C}}(h_{1},1 - h_{2};h_{2}/2 - h_{1})
    = \frac{1}{16}
    \frac{\Gamma(h_{1})^{4}\Gamma(h_{1}-\frac{h_{2}}{2})^{2}\Gamma(\frac{h_{2}}{2})^{4}}
    {\Gamma(h_{1}+\frac{h_{2}}{2})^{2}\Gamma(2h_{1}-h_{2})^{2}\Gamma(h_{2})^{2}}\,\mathcal{S},
\end{gather}
with $\mathcal{S}$ defined in \eqref{eq: leg integral 4}.
Combining the result with \eqref{eq: mixed vacuum three-point coefficient}, we find
\begin{gather}
    \label{eq: mixed thermal three-point result}
    \langle \! \ev{\mathcal{O}_{1}(0) \mathcal{O}_{1}(0) \mathcal{O}_{2}(0)} \! \rangle_{\beta}
    = -\frac{\kappa R^{3}}{32\pi^{2}}
    \left(\frac{2\pi}{\beta}\right)^{2\Delta_{1}+\Delta_{2}-4}\,\mathcal{A},
\end{gather}
where $\mathcal{A}$ is given in \eqref{eq: mixed correlator gravity 5}.
This precisely reproduces \eqref{eq: mixed correlator gravity 4} at $d=2$ and $z_{0}=\frac{\beta}{2\pi}$.

\section{Discussion}
\label{sec: discussion}
We have computed thermal two- and three-point correlators at vanishing external momenta in the black brane background.
The two-point function follows from the normalizable coefficient of the exact bulk-boundary propagator, while the contact three-point diagrams reduce to integrals of three Legendre functions.
These integrals give closed-form expressions both for identical scalar primaries and for the mixed correlator with two identical operators and a third of arbitrary dimension.
The results are exact in the operator dimensions and temperature at the bulk orders considered: the free two-point function and the tree-level three-point contributions.
In $d = 2$, we independently reproduced all three expressions by mapping vacuum CFT correlators to the thermal cylinder and evaluating their zero-momentum Fourier transforms as complex Selberg integrals. \\

We identified the poles arising from the trigonometric dependence of the zero-momentum thermal correlators on the operator dimension.
These poles are isolated on the positive real axis.
Analogous poles were seen in the thermal one-point function in the black brane background \cite{Grinberg:2020fdj,Berenstein:2022nlj}.\footnote{Thermal one-point functions in general backgrounds also exhibit a similar pole structure \cite{David:2022nfn}.}
Those poles were interpreted, in the semiclassical large-$N$ limit, as a consequence of enhanced operator mixing at resonant values of the operator dimension.
It would be interesting to develop a similar interpretation of the poles in higher-point thermal correlators. \\

By analytically continuing the conformal dimension to the lower half of the complex plane, we analyzed the large conformal dimension behaviour of the thermal correlators.
We showed that their asymptotic forms encode geometric information about the black hole, including a signature of its singularity.
More precisely, the leading exponential mass dependence is $e^{-2m\ell_{h}}$ for the two-point function and $e^{-3m\ell_{h}}$ for the three-point function, where $\ell_{h}$ is the regularized geodesic distance between the horizon and the asymptotic boundary.
The proper time to singularity, $\tau_{s}$, appears through powers of $e^{-im\tau_{s}}$ in the subleading exponential terms. \\

The exact three-point result for identical primaries disagrees with the leading form $e^{-3m(\ell_{h} + i\tau_{s})}$ predicted in \cite[equation (4.7)]{Singhi:2024sdr} for the continued zero-momentum contact contribution.
The earlier calculation retained the WKB prefactors and included a family of complex saddles near the singularity, but left the horizon endpoint contribution
uncontrolled because the WKB approximation breaks down there.
The discrepancy may arise from this endpoint contribution as well as from an incomplete account of the relevant saddles and their contributions.
The present calculation does not show how to repair the failure of the WKB calculation in the higher-point function, nor does it explain why the one-point saddle calculation succeeds.
However, the asymptotic expansion constrains how the proper time to the singularity enters: the leading exponential depends only on the boundary-to-horizon length, and terms of the form $e^{-3lim\tau_{s}}$ are absent for every positive integer $l$.
An exact WKB analysis \cite{kawai2005algebraic,Iwaki:2014vad} is a natural next step to address these questions.\footnote{An exact WKB approach was also used recently in \cite{Jia:2026ryl} to study the thermal spectral function and its connection to the black hole singularity.}

\acknowledgments
I am grateful to Justin R. David for important comments and prompt feedback on the original calculations.
I thank R. Loganayagam for several discussions on this work.
I also thank Saikat Ghosh, Ashoke Sen, Spenta Wadia, and the rest of the ICTS string group for various interactions during the course of this work.
I acknowledge support from the Department of Atomic Energy, Government of India, under project no. RTI4019.

\paragraph{Statement on AI use.} I used ChatGPT for some editing, figures, typesetting, and verification of certain claims in this work.
The manuscript was written by me and I take full responsibility for any errors.

\bibliographystyle{JHEP}
\bibliography{References}

@article{Maldacena:1997re,
    author = "Maldacena, Juan Martin",
    title = "{The Large $N$ limit of superconformal field theories and supergravity}",
    eprint = "hep-th/9711200",
    archivePrefix = "arXiv",
    reportNumber = "HUTP-97-A097, HUTP-98-A097",
    doi = "10.4310/ATMP.1998.v2.n2.a1",
    journal = "Adv. Theor. Math. Phys.",
    volume = "2",
    pages = "231--252",
    year = "1998"
}

@article{Gubser:1998bc,
    author = "Gubser, S. S. and Klebanov, Igor R. and Polyakov, Alexander M.",
    title = "{Gauge theory correlators from noncritical string theory}",
    eprint = "hep-th/9802109",
    archivePrefix = "arXiv",
    reportNumber = "PUPT-1767",
    doi = "10.1016/S0370-2693(98)00377-3",
    journal = "Phys. Lett. B",
    volume = "428",
    pages = "105--114",
    year = "1998"
}

@article{Witten:1998qj,
    author = "Witten, Edward",
    title = "{Anti de Sitter space and holography}",
    eprint = "hep-th/9802150",
    archivePrefix = "arXiv",
    reportNumber = "IASSNS-HEP-98-15",
    doi = "10.4310/ATMP.1998.v2.n2.a2",
    journal = "Adv. Theor. Math. Phys.",
    volume = "2",
    pages = "253--291",
    year = "1998"
}

@article{Singhi:2024sdr,
    author = "Singhi, Kaustubh",
    title = "{Proper time to the singularity and thermal correlators}",
    eprint = "2406.08553",
    archivePrefix = "arXiv",
    primaryClass = "hep-th",
    doi = "10.1103/n2n1-1d9f",
    journal = "Phys. Rev. D",
    volume = "112",
    number = "10",
    pages = "106011",
    year = "2025"
}

@article{Berenstein:2022nlj,
    author = "Berenstein, David and Mancilla, Robinson",
    title = "{Aspects of thermal one-point functions and response functions in AdS black holes}",
    eprint = "2211.05144",
    archivePrefix = "arXiv",
    primaryClass = "hep-th",
    doi = "10.1103/PhysRevD.107.126010",
    journal = "Phys. Rev. D",
    volume = "107",
    number = "12",
    pages = "126010",
    year = "2023"
}

@article{Grinberg:2020fdj,
    author = "Grinberg, Matan and Maldacena, Juan",
    title = "{Proper time to the black hole singularity from thermal one-point functions}",
    eprint = "2011.01004",
    archivePrefix = "arXiv",
    primaryClass = "hep-th",
    doi = "10.1007/JHEP03(2021)131",
    journal = "JHEP",
    volume = "03",
    pages = "131",
    year = "2021"
}

@article{David:2022nfn,
    author = "David, Justin R. and Kumar, Srijan",
    title = "{Thermal one point functions, large d and interior geometry of black holes}",
    eprint = "2212.07758",
    archivePrefix = "arXiv",
    primaryClass = "hep-th",
    doi = "10.1007/JHEP03(2023)256",
    journal = "JHEP",
    volume = "03",
    pages = "256",
    year = "2023"
}

@article{Rodriguez-Gomez:2021pfh,
    author = "Rodriguez-Gomez, D. and Russo, J. G.",
    title = "{Correlation functions in finite temperature CFT and black hole singularities}",
    eprint = "2102.11891",
    archivePrefix = "arXiv",
    primaryClass = "hep-th",
    reportNumber = "ICCUB-21-002",
    doi = "10.1007/JHEP06(2021)048",
    journal = "JHEP",
    volume = "06",
    pages = "048",
    year = "2021"
}

@article{Krishna:2021fus,
    author = "Krishna, Hare and Rodriguez-Gomez, D.",
    title = "{Holographic thermal correlators revisited}",
    eprint = "2108.00277",
    archivePrefix = "arXiv",
    primaryClass = "hep-th",
    doi = "10.1007/JHEP11(2021)139",
    journal = "JHEP",
    volume = "11",
    pages = "139",
    year = "2021"
}

@article{Kraus:2002iv,
    author = "Kraus, Per and Ooguri, Hirosi and Shenker, Stephen",
    title = "{Inside the horizon with AdS / CFT}",
    eprint = "hep-th/0212277",
    archivePrefix = "arXiv",
    reportNumber = "UCLA-02-TEP-41, CALT-68-2421, SU-ITP-02-45",
    doi = "10.1103/PhysRevD.67.124022",
    journal = "Phys. Rev. D",
    volume = "67",
    pages = "124022",
    year = "2003"
}

@article{Fidkowski:2003nf,
    author = "Fidkowski, Lukasz and Hubeny, Veronika and Kleban, Matthew and Shenker, Stephen",
    title = "{The Black hole singularity in AdS / CFT}",
    eprint = "hep-th/0306170",
    archivePrefix = "arXiv",
    reportNumber = "SU-ITP-03-16",
    doi = "10.1088/1126-6708/2004/02/014",
    journal = "JHEP",
    volume = "02",
    pages = "014",
    year = "2004"
}

@article{Festuccia:2005pi,
    author = "Festuccia, Guido and Liu, Hong",
    title = "{Excursions beyond the horizon: Black hole singularities in Yang-Mills theories. I.}",
    eprint = "hep-th/0506202",
    archivePrefix = "arXiv",
    reportNumber = "MIT-CTP-3641",
    doi = "10.1088/1126-6708/2006/04/044",
    journal = "JHEP",
    volume = "04",
    pages = "044",
    year = "2006"
}

@article{Ceplak:2024bja,
    author = "\v{C}eplak, Nejc and Liu, Hong and Parnachev, Andrei and Valach, Samuel",
    title = "{Black Hole Singularity from OPE}",
    eprint = "2404.17286",
    archivePrefix = "arXiv",
    doi = "10.1007/JHEP10(2024)105",
    journal = "JHEP",
    volume = "10",
    pages = "105",
    primaryClass = "hep-th",
    year = "2024"
}

@article{Jia:2026ryl,
    author = "Jia, Hewei Frederic and Rangamani, Mukund",
    title = "{Exact holographic thermal spectral functions: OPE, non-perturbative corrections, and black hole singularity}",
    eprint = "2604.10803",
    archivePrefix = "arXiv",
    primaryClass = "hep-th",
    month = "4",
    year = "2026"
}

@article{Afkhami-Jeddi:2025wra,
    author = "Afkhami-Jeddi, Nima and Caron-Huot, Simon and Chakravarty, Joydeep and Maloney, Alexander",
    title = "{Imprint of the black hole singularity on thermal two-point functions}",
    eprint = "2510.21673",
    archivePrefix = "arXiv",
    primaryClass = "hep-th",
    doi = "10.1103/1j3s-fzjl",
    journal = "Phys. Rev. D",
    volume = "114",
    number = "2",
    pages = "025005",
    year = "2026"
}

@article{Giombi:2026kdz,
    author = "Giombi, Simone and Li, Yue-Zhou and Shan, Jieru",
    title = "{Bouncing singularities and thermal correlators on line defects}",
    eprint = "2603.11012",
    archivePrefix = "arXiv",
    primaryClass = "hep-th",
    doi = "10.1007/JHEP07(2026)170",
    journal = "JHEP",
    volume = "07",
    pages = "170",
    year = "2026"
}

@article{Festuccia:2008zx,
    author = "Festuccia, Guido and Liu, Hong",
    title = "{A Bohr-Sommerfeld quantization formula for quasinormal frequencies of AdS black holes}",
    eprint = "0811.1033",
    archivePrefix = "arXiv",
    primaryClass = "gr-qc",
    doi = "10.1166/asl.2009.1029",
    journal = "Adv. Sci. Lett.",
    volume = "2",
    number = "2",
    pages = "221--235",
    year = "2009"
}

@article{Arnaudo:2026tcy,
    author = "Arnaudo, Paolo and Withers, Benjamin",
    title = "{Bouncing singularities in Schwarzschild: a geometric origin of the QNM convergence region}",
    eprint = "2605.16489",
    archivePrefix = "arXiv",
    primaryClass = "gr-qc",
    month = "5",
    year = "2026"
}

@article{Horowitz:2023ury,
    author = "Horowitz, Gary T. and Leung, Henry and Queimada, Leonel and Zhao, Ying",
    title = "{Boundary signature of singularity in the presence of a shock wave}",
    eprint = "2310.03076",
    archivePrefix = "arXiv",
    primaryClass = "hep-th",
    doi = "10.21468/SciPostPhys.16.2.060",
    journal = "SciPost Phys.",
    volume = "16",
    pages = "060",
    year = "2024"
}

@article{Ceplak:2025dds,
    author = "\v{C}eplak, Nejc and Liu, Hong and Parnachev, Andrei and Valach, Samuel",
    title = "{Fooling the censor: going beyond inner horizons with the OPE}",
    eprint = "2511.09638",
    archivePrefix = "arXiv",
    primaryClass = "hep-th",
    doi = "10.1007/JHEP05(2026)001",
    journal = "JHEP",
    volume = "05",
    pages = "001",
    year = "2026"
}

@article{AliAhmad:2026wem,
    author = "Ali Ahmad, Shadi and Almheiri, Ahmed and Lin, Simon",
    title = "{Continuing past the inner horizon using WKB}",
    eprint = "2601.02354",
    archivePrefix = "arXiv",
    primaryClass = "hep-th",
    month = "1",
    year = "2026"
}

@article{Chakravarty:2026fdm,
    author = "Chakravarty, Joydeep and Zigdon, Yoav",
    title = "{A signature of the timelike singularity in holography}",
    eprint = "2610.30134",
    archivePrefix = "arXiv",
    primaryClass = "hep-th",
    month = "10",
    year = "2026"
}

@article{Chakravarty:2025ncy,
  author = {Chakravarty, Joydeep},
  title = {{Imprint of black hole interior on thermal four-point correlators}},
  eprint = {2512.10912},
  archiveprefix = {arXiv},
  primaryclass = {hep-th},
  doi = {10.1007/JHEP09(2026)034},
  journal = {JHEP},
  volume = {09},
  pages = {034},
  year = {2026},
}

@article{Dodelson:2025jff,
    author = "Dodelson, Matthew and Iossa, Cristoforo and Karlsson, Robin",
    title = "{Bouncing off a stringy singularity}",
    eprint = "2511.09616",
    archivePrefix = "arXiv",
    primaryClass = "hep-th",
    doi = "10.1007/JHEP07(2026)161",
    journal = "JHEP",
    volume = "07",
    pages = "161",
    year = "2026"
}

@article{Buric:2026qsp,
    author = "Buri{\'c}, Ilija and Chang, Chi-Ming and Gusev, Ivan and Helfenberger, Elizabeth and Parnachev, Andrei and Rangamani, Mukund",
    title = "{Thermal two-point functions in SYK and complex-time singularities}",
    eprint = "2607.05258",
    archivePrefix = "arXiv",
    primaryClass = "hep-th",
    month = "7",
    year = "2026"
}

@article{Buric:2026sit,
    author = "Buri{\'c}, Ilija and Gusev, Ivan and Parnachev, Andrei",
    title = "{Stringy resolution of bouncing singularities}",
    eprint = "2610.12311",
    archivePrefix = "arXiv",
    primaryClass = "hep-th",
    month = "10",
    year = "2026"
}

@article{Komargodski:2026oob,
    author = "Komargodski, Zohar",
    title = "{Fate of Bouncing Singularities at Finite $G_N$}",
    eprint = "2609.31309",
    archivePrefix = "arXiv",
    primaryClass = "hep-th",
    month = "9",
    year = "2026"
}

@article{Batra:2026gbq,
    author = "Batra, Gauri and Levine, Adam and Shenker, Stephen H.",
    title = "{A backreacted bouncing geodesic in 2D}",
    eprint = "2609.32071",
    archivePrefix = "arXiv",
    primaryClass = "hep-th",
    month = "9",
    year = "2026"
}

@article{Liu:2026Smoothing,
    author = "Liu, Mingxuan and Ruan, Shan-Ming",
    title = "{Smoothing the Bouncing-Geodesic Singularity in AdS/CFT}",
    eprint = "2611.03347",
    archivePrefix = "arXiv",
    primaryClass = "hep-th",
    month = "11",
    year = "2026"
}

@article{Bzowski:2013sza,
    author = "Bzowski, Adam and McFadden, Paul and Skenderis, Kostas",
    title = "{Implications of conformal invariance in momentum space}",
    eprint = "1304.7760",
    archivePrefix = "arXiv",
    primaryClass = "hep-th",
    doi = "10.1007/JHEP03(2014)111",
    journal = "JHEP",
    volume = "03",
    pages = "111",
    year = "2014"
}

@article{Yarom:2009mw,
    author = "Yarom, Amos",
    title = "{Notes on the bulk viscosity of holographic gauge theory plasmas}",
    eprint = "0912.2100",
    archivePrefix = "arXiv",
    primaryClass = "hep-th",
    reportNumber = "PUPT-2322",
    doi = "10.1007/JHEP04(2010)024",
    journal = "JHEP",
    volume = "04",
    pages = "024",
    year = "2010"
}

@article{Freedman:1998tz,
    author = "Freedman, Daniel Z. and Mathur, Samir D. and Matusis, Alec and Rastelli, Leonardo",
    title = "{Correlation functions in the CFT(d) / AdS(d+1) correspondence}",
    eprint = "hep-th/9804058",
    archivePrefix = "arXiv",
    reportNumber = "MIT-CTP-2727",
    doi = "10.1016/S0550-3213(99)00053-X",
    journal = "Nucl. Phys. B",
    volume = "546",
    pages = "96--118",
    year = "1999"
}

@article{Son:2002sd,
    author = "Son, Dam T. and Starinets, Andrei O.",
    title = "{Minkowski space correlators in AdS / CFT correspondence: Recipe and applications}",
    eprint = "hep-th/0205051",
    archivePrefix = "arXiv",
    reportNumber = "INT-PUB-02-34",
    doi = "10.1088/1126-6708/2002/09/042",
    journal = "JHEP",
    volume = "09",
    pages = "042",
    year = "2002"
}

@article{Skenderis:2002wp,
  author = {Skenderis, Kostas},
  editor = {de Wit, B. and Vandoren, S.},
  title = {{Lecture notes on holographic renormalization}},
  eprint = {hep-th/0209067},
  archiveprefix = {arXiv},
  reportnumber = {PUTP-2047},
  doi = {10.1088/0264-9381/19/22/306},
  journal = {Class. Quant. Grav.},
  volume = {19},
  pages = {5849--5876},
  year = {2002},
}

@misc{NIST:DLMF,
         key = "{\relax DLMF}",
       title = "{\it NIST Digital Library of Mathematical Functions}",
howpublished = "\url{https://dlmf.nist.gov/}, Release 1.2.4 of 2025-03-15",
         url = "https://dlmf.nist.gov/",
        note = "F.~W.~J. Olver, A.~B. {Olde Daalhuis}, D.~W. Lozier, B.~I. Schneider,
                R.~F. Boisvert, C.~W. Clark, B.~R. Miller, B.~V. Saunders,
                H.~S. Cohl, and M.~A. McClain, eds."}

@article{Bros:2009bz,
    author = "Bros, Jacques and Epstein, Henri and Gaudin, Michel and Moschella, Ugo and Pasquier, Vincent",
    title = "{Triangular invariants, three-point functions and particle stability on the de Sitter universe}",
    eprint = "0901.4223",
    archivePrefix = "arXiv",
    primaryClass = "hep-th",
    doi = "10.1007/s00220-009-0875-4",
    journal = "Commun. Math. Phys.",
    volume = "295",
    pages = "261--288",
    year = "2010"
}

@article{Breitenlohner:1982bm,
    author = {{Breitenlohner}, Peter and {Freedman}, Daniel Z.},
    title = "{Positive energy in anti-de Sitter backgrounds and gauged extended supergravity}",
    journal = {Physics Letters B},
    year = "1982",
    volume = "115",
    number = "3",
    pages = "197--201",
    doi = "10.1016/0370-2693(82)90643-8"
}

@article{Becker:2014jla,
  author = {Becker, Melanie and Cabrera, Yaniel and Su, Ning},
  title = {{Finite-temperature three-point function in 2D CFT}},
  eprint = {1407.3415},
  archiveprefix = {arXiv},
  primaryclass = {hep-th},
  reportnumber = {MIFPA-14-21},
  doi = {10.1007/JHEP09(2014)157},
  journal = {JHEP},
  volume = {09},
  pages = {157},
  year = {2014},
}

@article{Aomoto:1987,
    author = "Aomoto, Kazuhiko",
    title = "{On the complex Selberg integral}",
    journal = "Quart. J. Math. Oxford Ser. (2)",
    volume = "38",
    pages = "385--399",
    year = "1987"
}

@article{Dotsenko:1984nm,
    author = "Dotsenko, Vl. S. and Fateev, V. A.",
    title = "{Four-point correlation functions and the operator algebra in 2D conformal invariant theories with central charge $C \leq 1$}",
    journal = "Nucl. Phys. B",
    volume = "251",
    pages = "691--734",
    year = "1985"
}

@article{Forrester:2007,
    author = "Forrester, Peter J. and Warnaar, S. Ole",
    title = "{The importance of the Selberg integral}",
    eprint = "0710.3981",
    archivePrefix = "arXiv",
    primaryClass = "math.CA",
    journal = "Bull. Amer. Math. Soc. (N.S.)",
    volume = "45",
    pages = "489--534",
    year = "2008"
}

@article{Neretin:2022pbv,
    author = "Neretin, Yury A.",
    title = "{On the Dotsenko--Fateev complex twin of the Selberg integral and its extensions}",
    eprint = "2212.09112",
    archivePrefix = "arXiv",
    primaryClass = "math.CA",
    doi = "10.1007/s11139-023-00804-3",
    journal = "Ramanujan J.",
    year = "2023"
}

@book{kawai2005algebraic,
  title={Algebraic Analysis of Singular Perturbation Theory},
  author={Kawai, T. and Takei, Y.},
  isbn={9780821835470},
  lccn={2005048161},
  series={Iwanami series in modern mathematics},
  url={https://books.google.co.in/books?id=GDsjJLinxogC},
  year={2005},
  publisher={American Mathematical Society}
}

@article{Iwaki:2014vad,
    author = "Iwaki, Kohei and Nakanishi, Tomoki",
    title = "{Exact WKB analysis and cluster algebras}",
    eprint = "1401.7094",
    archivePrefix = "arXiv",
    primaryClass = "math.CA",
    doi = "10.1088/1751-8113/47/47/474009",
    journal = "J. Phys. A",
    volume = "47",
    number = "47",
    pages = "474009",
    year = "2014"
}

\end{document}